\documentclass[twocolumn]{revtex4-1}
\usepackage{xcolor}
\usepackage{graphicx}
\usepackage{amsmath}

\usepackage{mathtext}
\usepackage{braket}

\usepackage[colorlinks=true, linkcolor=cyan, urlcolor=blue, citecolor=blue]{hyperref}

\begin{document}

\title{Quantification of electron correlation for approximate quantum calculations} 
\author{Shunyue Yuan}
\affiliation{Department of Physics, University of Illinois at Urbana-Champaign} 
\author{Yueqing Chang}
\affiliation{Department of Physics, University of Illinois at Urbana-Champaign} 
\author{Lucas K. Wagner}
\affiliation{Department of Physics, University of Illinois at Urbana-Champaign} 
\begin{abstract}
State-of-the-art many-body wave function techniques rely on heuristics to achieve high accuracy at an attainable cost to solve the many-body Schr\"odinger equation. 
By far the most common property used to assess accuracy has been the total energy; however, total energies do not give a complete picture of electron correlation.. 
In this work, the authors assess the von Neumann entropy of the one-particle reduced density matrix (1-RDM)  to compare selected configuration interaction (CI), coupled cluster, variational Monte Carlo, and fixed-node diffusion Monte Carlo for benchmark hydrogen chains.
A new algorithm, the circle reject method is presented which improves the efficiency of the evaluation of the von Neumann entropy using quantum Monte Carlo by several orders of magnitude. 
The von Neumann entropy of the 1-RDM and the eigenvalues of the 1-RDM are shown to distinguish between the dynamic correlation introduced by the Jastrow and static correlation introduced by determinants with large weights, confirming some of the lore in the field concerning the difference between the selected CI and Slater-Jastrow wave functions.
\end{abstract}
\maketitle

\section*{Introduction}

The development of computational algorithms to solve the many-electron problems is one of the grand challenges in modern physics, chemistry, and materials science. 
Such algorithms allow for accurate simulation of essentially all of chemistry and materials science, and indeed a significant fraction of computer time is devoted to these simulations.
Currently, density functional theory (DFT) is by far the most common technique to achieve this goal; however, because of the unknown functional, it is difficult to systematically improve the performance despite significant attempts \cite{sun_strongly_2015, sun_accurate_2016, morales_quantum_2014, borlido_large-scale_2019, borlido_exchange-correlation_2020}.
Wave function techniques, such as quantum Monte Carlo (QMC) \cite{ ceperley_ground_1980, Foulkes}, coupled cluster (CC) \cite{cizek_coupled_1980, paldus_coupled_1989, bartlett_coupled-cluster_2007}, density matrix renormalization group (DMRG) \cite{whiteDensityMatrixFormulation1992, schollwock_density-matrix_2005}, or various truncated configuration interaction (CI) methods \cite{huron_iterative_1973, cimiraglia_recent_1987, HCI}, offer a systematically improvable path to accurate quantum simulations, at the cost of larger computational expense compared with mean-field theories. 
Na\"ive methods such as exact diagonalization scale exponentially in general; high accuracy at an attainable computational cost is only obtained by using heuristics.
For example, fixed-node diffusion Monte Carlo requires accurate wave function nodes, CC uses an exponential \textit{ansatz}, and CI methods must select the determinants to include.

It is interesting to compare the heuristic nature of many-electron algorithms to the no free lunch theorem \cite{wolpertNoFreeLunch1997} in optimization. 
Shortly stated, any two optimization algorithms are equivalent in performance when averaged across all possible problems. 
However, in practice some optimization algorithms perform much better than others on problems in a given class. 
In many-electron simulation, we are concerned with problems that represent realistic physical situations, which is a very small subclass of all problems. 
While some many-body problems are proveably computationally hard \cite{troyerComputationalComplexityFundamental2005}, it is not always clear \textit{a priori} which heuristics will lead to accurate and efficient solutions, and how to assess different heuristics in a way that allows insight into how they treat electron correlation. 
The current state of the art focuses on total energy comparisons \cite{simonscollaborationonthemany-electronproblemDirectComparisonManyBody2020}, which, while important, often does not offer much insight into how the choice of approach affects the treatment of electron correlation. 

There exist a number of approaches to quantify electronic correlation, each with their advantages and disadvantages. 
For example, the spacial entanglement \cite{amico_entanglement_2008} has a close relationship with the performance of density matrix renormalization group \cite{vidal_efficient_2003, vidal_efficient_2004}; however, it requires the replica trick in Monte Carlo \cite{mcminisRenyiEntropyInteracting2013}, which can be rather expensive computationally.
Similarly, the two-particle reduced density matrix (2-RDM) is often too expensive to compute in its entirety, so measures such as the cumulant two-particle reduced density matrix \cite{juhaszCumulantTwoparticleReduced2006b} can be impractical for larger calculations.
Other proposed measures \cite{dellesiteShannonEntropyManyelectron2015a,ramos-cordobaSeparationDynamicNondynamic2016} rely on the definition of a particular reference, which we did not find suitable for benchmarking across multiple methods.
Finally, we should mention the idea of orbital-based entanglement measures \cite{boguslawskiEntanglementMeasuresSingle2012} which are well suited for understanding density matrix renormalization group \cite{whiteDensityMatrixFormulation1992} performance, but again require the 2-RDM.
To compare disparate methods that may be under active development, it is critical that a quantification of correlation is very simple to evaluate.

In this work, we assess multipartite entanglement, defined as the von Neumann entropy of the one-particle reduced density matrix, as a tool to understand standard heuristics for treating electron correlation in many-particle wave functions, and test it versus very different approaches to including electron correlation. 
We develop a new technique based on rejection of eigenvalues that improves the performance of Monte Carlo evaluations of von Neumann entropy by several orders of magnitude.
We find that the multipartite entanglement quantifies much of current lore about how different wave function \textit{ansatzes} add correlations.
For example, Jastrow correlation factors are often said \cite{filippiMulticonfigurationWaveFunctions1996} to capture dynamic correlation, while configuration interaction with a few determinants captures static correlation. 
We use the von Neumann entropy of the one-particle reduced density matrix (1-RDM) to characterize electron correlation, and find that the entropy of the 1-RDM correlates closely with these ideas. 

\section*{Multipartite entanglement}

We quantify the multipartite entanglement of a many-electron wave function $\Psi$ using its 1-RDM
\begin{equation}
  \rho_{ij,\sigma} = \langle\Psi|c_{i,\sigma}^\dagger c_{j,\sigma}|\Psi\rangle,
\end{equation}
where $c^\dagger_{i\sigma}$ and $c_{i\sigma}$ are creation and annihilation operators for the single-particle orbital $\phi_i$ with spin $\sigma$.
We define the entanglement entropy as
\begin{equation}
    S = -\mathrm{Tr}\left( \rho \ln \rho\right).
\end{equation}
One can rewrite this using the entanglement spectrum, i.e., the eigenvalues of the 1-RDM, $\lambda_i$, as
\begin{equation}
    S = -\sum_i \lambda_i \ln{\lambda_i}.
    \label{eqn:entropy_using_spectrum}
\end{equation}

The multipartite entanglement entropy measures how much information is lost when a single determinant is used to describe the wave function. 
It is monotonically related to the quasiparticle renormalization factor that appears in Fermi liquid theory, which also can be computed in quantum Monte Carlo as a measure of correlation \cite{holzmannMomentumDistributionHomogeneous2011b}. 
As we shall show in this paper, the spectrum of the 1-RDM appears to give extra information about the type of correlation present in the wave function.


\section*{Methods}

\subsection{Electronic structure methods}
\label{sec:electronic_structure_methods}
\begin{figure}
    \centering
    \includegraphics{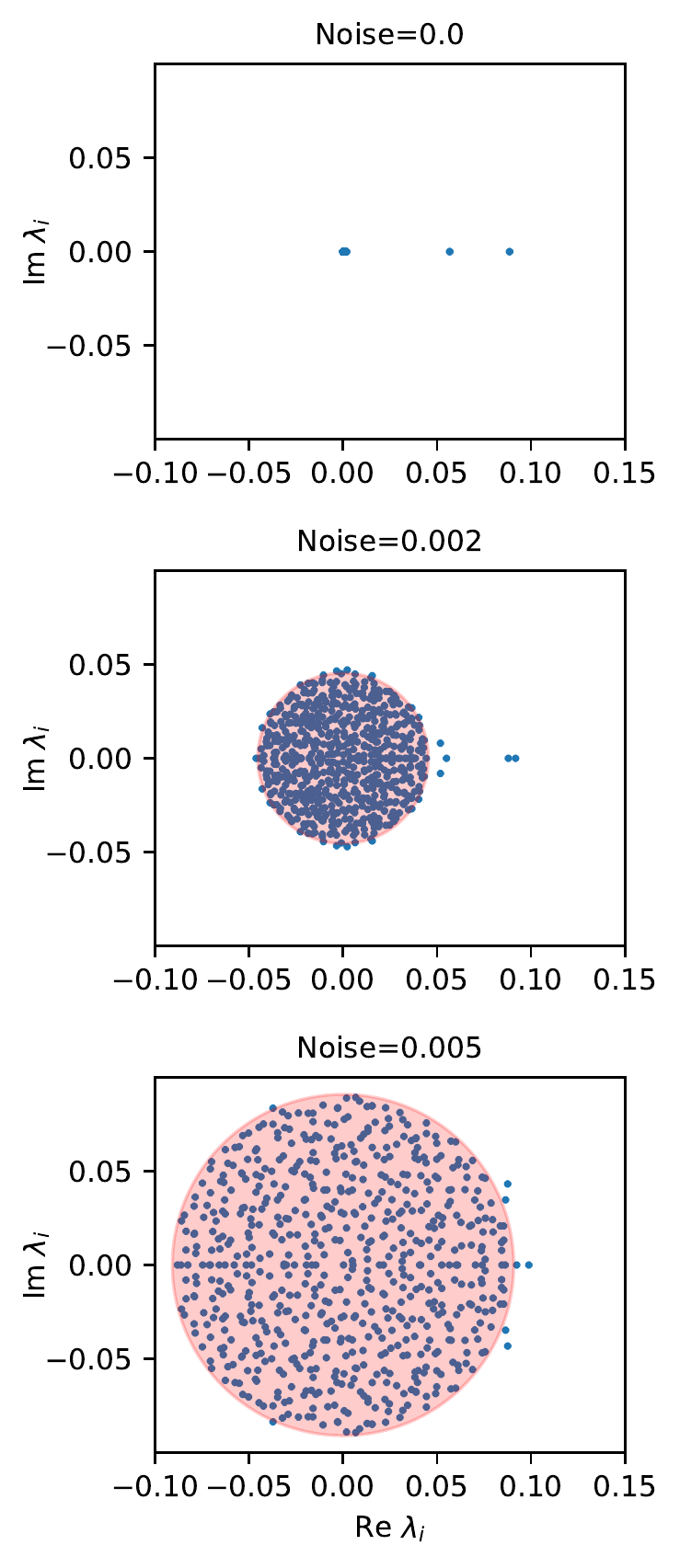}
    \caption{The eigenvalues $\lambda_i$ of the 1-RDM with added noise $\sigma$ follow the circular distribution.} 
    \label{fig:eigenvalue_distribution}
\end{figure}

In this work, we perform calculations on one dimensional chains of $N$ equally spaced hydrogen atoms, where $N = 2, 4, 6, 8, 10$.
We consider systems with interatomic separation $r$ equals to $1.4$ and $3.0$, in units of the Bohr radius ($a_{\text{B}} = \hbar^2/(me^2)$) to compare weak and strong correlations. 
The ground-state wave functions are generated using Hartree-Fock (HF), heat-bath configuration interaction (HCI), coupled cluster with singles and doubles (CCSD), variational Monte Carlo (VMC), and fixed-node diffusion Monte Carlo (FN-DMC).
CCSD in H6 and smaller systems are performed using a correlation consistent 5-zeta valence basis set (cc-pV5Z), and for all the other calculations we always use correlation consistent triple-zeta valence basis set (cc-pVTZ) \cite{Basis}.

We start from constructing HF and HCI wave functions. 
Each HCI wave function is specified by the threshold of the Hamiltonian matrix element $\epsilon_1$, which controls which determinants are included in the wave function \cite{HCI}.
We gradually decrease $\epsilon_1$ until the energy converges. 
Due to limited computational resources, converged HCI wave functions are obtained only for hydrogen chains with $N\leq 6$.

We use multi-determinant Slater-Jastrow (MSJ) wave functions to perform VMC and FN-DMC calculations. 
In the special case of a single determinant, we will refer to the wavefunction as simply Slater-Jastrow (SJ). 
The SJ wave functions are constructed by multiplying the HF wave function of the lowest $n$ molecular orbitals $\lbrace \phi_i\rbrace$ by a two-body Jastrow factor $e^U$ \cite{Jastrow}, 
\begin{equation}
\Psi_{\text{SJ}} = e^{U} D^{\uparrow}\left[\phi_i(\mathbf{r}_j)\right] D^{\downarrow}\left[\phi_i(\mathbf{r}_j)\right].
\label{eqn:SJ}
\end{equation}
The MSJ wave functions are constructed as
\begin{equation}
\Psi_{\text{MSJ}} = e^{U} \sum_{|c_\alpha|\leq \epsilon_2} c_{\alpha} D_{\alpha}^{\uparrow}\left[\phi_i(\mathbf{r}_j)\right] D_{\alpha}^{\downarrow}\left[\phi_i(\mathbf{r}_j)\right], \ 
\label{eqn:MSJ}
\end{equation}
where the determinants are taken from an HCI calculation with cutoff $\epsilon_1$, and further selected by including only determinants with coefficient $|c_\alpha| < \epsilon_2$.
Starting from the HCI wave function, we use VMC to optimize the parameters in the Jastrow factor $U$, the molecular orbitals $\lbrace\phi_i\rbrace$ and the determinant coefficients $\lbrace c_{\alpha}\rbrace$, then apply FN-DMC to project out the ground state at a time step of $\tau = 0.02$.

Since the operator $\hat{\rho}$ does not commute with the Hamiltonian, we use the extrapolated estimator \cite{Foulkes} to evaluate the 1-RDM of the fixed-node wave function $\Psi_{\text{FN}}$,
\begin{equation}
    \rho_{\text{extrapolated}} = \rho_{\text{mixed}} + \rho^\dagger_{\text{mixed}} - \rho_{\text{VMC}},
\label{eqn:extrapolated_estimator}
\end{equation}
where $\rho_{\text{mixed}}=\braket{\Psi_{\text{FN}}|\hat{\rho}|\Psi_{\text{T}}}$ and $\rho_{\text{VMC}}=\braket{\Psi_{\text{T}}|\hat{\rho}|\Psi_{\text{T}}}$. 
Here, $\ket{\Psi_{\text{T}}}$ and $\ket{\Psi_{\text{FN}}}$ are the optimized trial wave function and fixed-node wave function.
To derive equation~\ref{eqn:extrapolated_estimator}, we take $\delta \Psi = \Psi_{\text{FN}} - \Psi_{\text{T}}$, and only keep $\mathcal{O}\left(\delta\Psi\right)$,
\begin{align}
\begin{split}
    \braket{\Psi_{\text{FN}}|\hat{\rho}|\Psi_{\text{FN}}} &= \braket{\left(\Psi_{\text{T}}+\delta\Psi\right)|\hat{\rho}|\left(\Psi_{\text{T}}+\delta\Psi\right)} \\
    & \approx \braket{\Psi_{\text{T}}|\hat{\rho}|\Psi_{\text{T}}} + \braket{\delta\Psi |\hat{\rho}|\Psi_{\text{T}}} + \braket{\Psi_{\text{T}} |\hat{\rho}|\delta\Psi}   \\
    &= \braket{\Psi_{\text{FN}}|\hat{\rho}|\Psi_{\text{T}}} + \braket{\Psi_{\text{T}}|\hat{\rho}|\Psi_{\text{FN}}} - \braket{\Psi_{\text{T}}|\hat{\rho}|\Psi_{\text{T}}}. 
\end{split}
\notag
\label{eqn:derivation}
\end{align}
It is important that we evaluate $\rho$ and $\rho^\dagger$ using separate FN-DMC calculations in order for the circular distribution to be obeyed, and thus for the circle reject algorithm to be applicable. 
If $\rho$ and $\rho^\dagger$ are the same stochastic evaluation of the density matrix, then the resulting extrapolated density matrix is a mixture of symmetrized and non-symmetrized random matrices, which as we showed in Section~\ref{sec:circle_reject} leads to a large bias in the estimated entropy.

All quantum Monte Carlo (QMC) calculations are performed using \texttt{PyQMC} package \cite{pyqmc}, and the HF, HCI, and CCSD calculations are done using the \texttt{PySCF} package \cite{pyscf}. 
We perform these calculations using a \texttt{snakemake} workflow, openly available in the GitHub repository ``Energy-Entropy", reference number \cite{Workflow}. 

\subsection{Computing the entanglement entropy for stochastic matrices: circle reject algorithm}
\label{sec:circle_reject}
\begin{figure}
    \centering
    \includegraphics{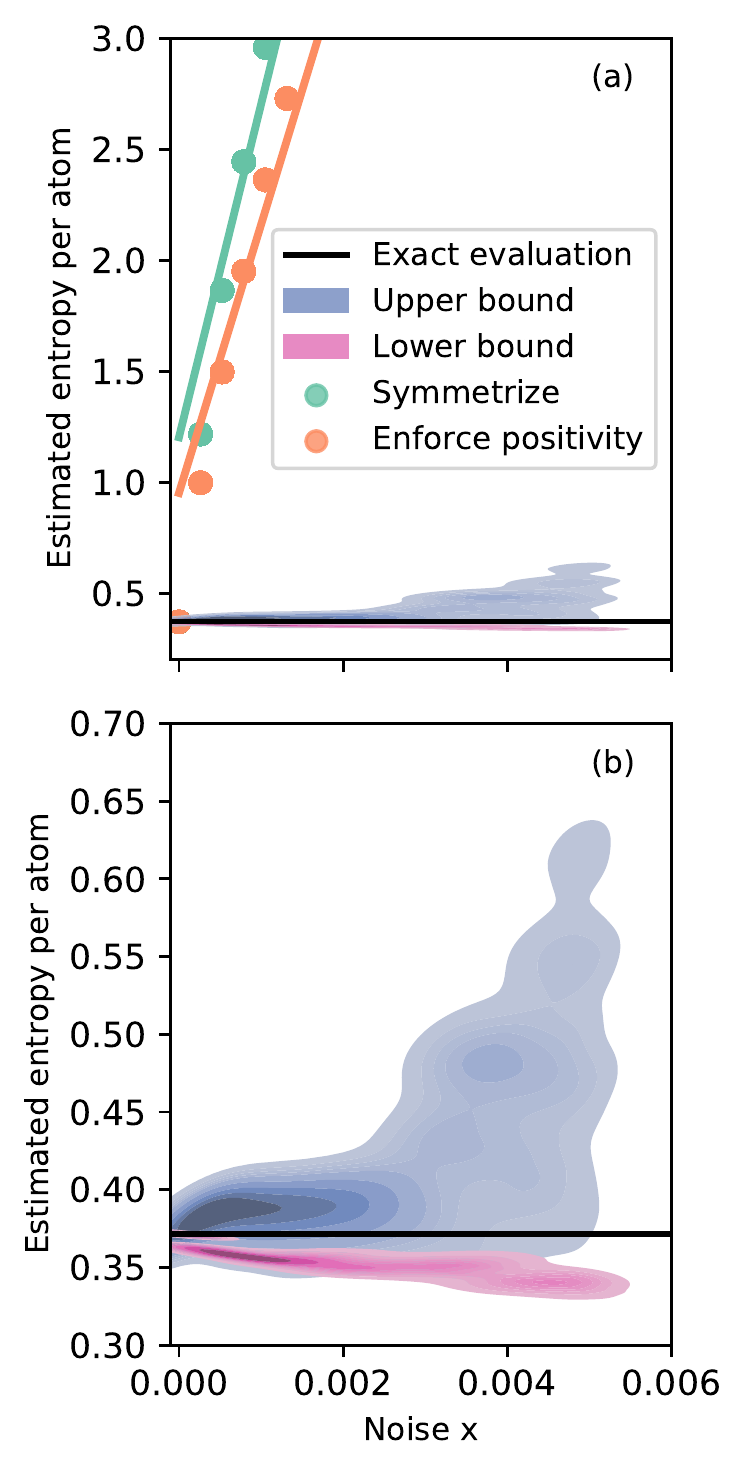}
    \caption{(a) Estimated entropy per atom using different strategies versus the noise added to a CCSD 1-RDM. The circle reject technique corrects bias in the entropy due to statistical fluctuations. The upper and lower bounds of entropy are estimated from the missing trace and bound above (below) the exact result. (b) Zoomed-in circle reject estimations shown in (a).}
    \label{fig:correction_schemes}
\end{figure}

The entanglement entropy has a bias when evaluated na\"ively on a matrix with stochastic noise using equation~(\ref{eqn:entropy_using_spectrum}).
So it is necessary to develop a method to compute the entropy correctly from quantum Monte Carlo evaluations of the 1-RDM. 
Assume the true value of the 1-RDM is $\bar{A}$ and its quantum Monte Carlo evaluation is $A$ with uncertainty $\epsilon$, then the probability density of $A$ is 
\begin{equation}
    \rho(A|\bar{A},\epsilon) \propto \prod_{ij} \exp\left[ \frac{(A_{ij}-\bar{A}_{ij})^2}{2\epsilon_{ij}^2} \right].
\end{equation}
We would like to evaluate the true von Neumann entropy 
\begin{equation}
\bar{s} = - \sum_i \bar{\lambda}_i \ln \bar{\lambda}_i,
\end{equation}
where $\bar{\lambda}_i$ are the eigenvalues of the matrix $\bar{A}$.

Our objective is to infer, from $A$, the most probable values of $\bar{\lambda}_i$, and therefore $\bar{s}$.
One complication is that most physical 1-RDMs have only a few non-zero eigenvalues; most are close to 1 or 0. 
In contrast, a random positive definite matrix has eigenvalues which almost always deviate from 0 and 1. 
For the part of the RDM that is zero, the stochastic noises brought by QMC distribute eigenvalues uniformly in a circle \cite{Circular_Law}. 
This gives rise to a bias in the computed entropy. 

To illustrate the distribution of eigenvalues of noisy matrices, we add Gaussian noise with standard deviation $\sigma$ to each element of the 1-RDM $A_{ij} = \bar{A}_{ij}+\chi_{ij}, \ \chi_{ij}\sim \mathcal{N}(0, \sigma)$,  where $\bar{A}$ is computed using CCSD for H6 at a 3.0 $a_{\text{B}}$ separation and a cc-pV5Z basis set. 
Fig.~\ref{fig:eigenvalue_distribution} shows the eigenvalues of the matrix $A$ with standard deviation $\sigma = 0.0, 0.002$ and $0.005$.
The circular distributions of the eigenvalues are highlighted by the red circles on the plot, with radii determined by $\sigma\sqrt{N}$, where $N$ is the size of the matrix.
The noise in the eigenvalue spectrum is covered by the circle given by random matrix theory \cite{Circular_Law}.

We consider three strategies of reducing the bias in entropy na\"ively computed using equation~\ref{eqn:entropy_using_spectrum} due to the presence of the noise, shown in Fig.~\ref{fig:correction_schemes} (a).
\begin{enumerate} 
\item \textbf{Symmetrize the matrix} by diagonalizing $\frac{A+A^\dagger}{2}$.
\item \textbf{Enforce positivity} by diagonalizing $A$ and setting all  negative eigenvalues to zero, and all imaginary components to zero.
\item \textbf{Circle reject} by removing all eigenvalues within the circular distribution given in red in Fig~\ref{fig:eigenvalue_distribution}.
\end{enumerate}
In Fig~\ref{fig:correction_schemes}, we show the bias in the entropy as a function of the noise $\sigma$ added to a CCSD 1-RDM. 
It is clear from the figure that the circle reject algorithm (noted by the upper and lower bounds) has a dramatically lower bias than the other strategies.

Our best strategy is the circle reject algorithm, which we give in detail here.
The strategy is as follows:
\begin{enumerate}
    \item Compute the eigenvalues $\lambda_i$ of matrix $A$. 
    \item Estimate the radius $r= \sigma\sqrt{N}$ of the circle.
    \item Adjust the radius as 
    
    $r' = \max \left( \left\lbrace|\lambda_i| : |\lambda_i|>r, ~|\operatorname{Re} (\lambda_i)|<r\right\rbrace \right) + \delta$
    \item Compute the first estimate of entropy as
    
    $S = -\sum_{|\lambda_i| \geq r'} \lambda_i \ln(\lambda_i)$.
    \item Estimate the upper and lower bounds of entropy using Eqn~\ref{eqn:upper_bound} and Eqn~\ref{eqn:lower_bound}.
\end{enumerate}

We found that step 3 improved the performance of the algorithm, since occasionally the noise falsely brings some small eigenvalues out of the circle with radius $r$.
Step 3 makes sure these eigenvalues are rejected.

In step 5, we estimate the upper and lower bounds by distributing the missing trace due to rejection in different ways. 
The total trace of 1-RDM should be $N$, where $N$ is the number of electrons. 
For the lower bound of the entropy, we equally re-distribute the missing trace among $m$ eigenvalues such that they are just below the rejection radius $r'$, i.e.,
\begin{align}
\begin{split}
  &\lambda_{\text{l}} = \frac{N-\sum_{|\lambda_i| \geq r'} \lambda_i}{m}, \\
  &m = \left\lceil \frac{N-\sum_{|\lambda_i| \geq r'} \lambda_i}{r'} \right\rceil,
\end{split}
\label{eqn:eigenvalues_lower}
\end{align}
where $\lceil \rceil$ indicates the smallest integer greater than the argument,
 and $\lambda_i$ are the eigenvalues of 1-RDM.
We compute the lower bound as
 \begin{equation}
 S_{\text{l}} = -\sum_{\left|\lambda_{i}\right| \geq r'} \lambda_{i} \ln \left(\lambda_{i}\right)
 - m \lambda_{\text{l}} \ln \lambda_{\text{l}}.
 \label{eqn:upper_bound}
\end{equation}
To estimate the upper bound of the entropy, we equally re-distribute the missing trace among all the missing eigenvalues, such that
\begin{equation}
    \lambda_{\text{u}}=\frac{N-\sum_{|\lambda_i| \geq r'} \lambda_i}{n},
    \label{eqn:eigenvalues_upper}
\end{equation}
 where $n$ is the total number of eigenvalues that are rejected.
\begin{equation}
    S_{\text{u}} = 
-\sum_{\left|\lambda_{i}\right| \geq r'} \lambda_{i} \ln \left(\lambda_{i}\right)
 - n \lambda_{\text{u}} \ln \lambda_{\text{u}}.
 \label{eqn:lower_bound}
 \end{equation}

Fig.~\ref{fig:correction_schemes} shows that the circle reject algorithm is much more efficient for this type of matrix than the other two strategies we considered. 
The lower bound derived is always a strict lower bound, but the upper bound occasionally falls below the true value.
The upper bound fails when a statistical fluctuation results in an enhancement of the eigenvalues outside the circle reject radius, so that the maximum missing entropy is underestimated. 
For the rest of the paper, the estimated upper and lower bounds will be reported in all estimations of the entropy using stochastic methods (VMC and DMC), in lieu of single-$\sigma$ uncertainties. 

\section*{Results} 

\begin{figure}
    \centering
    \includegraphics{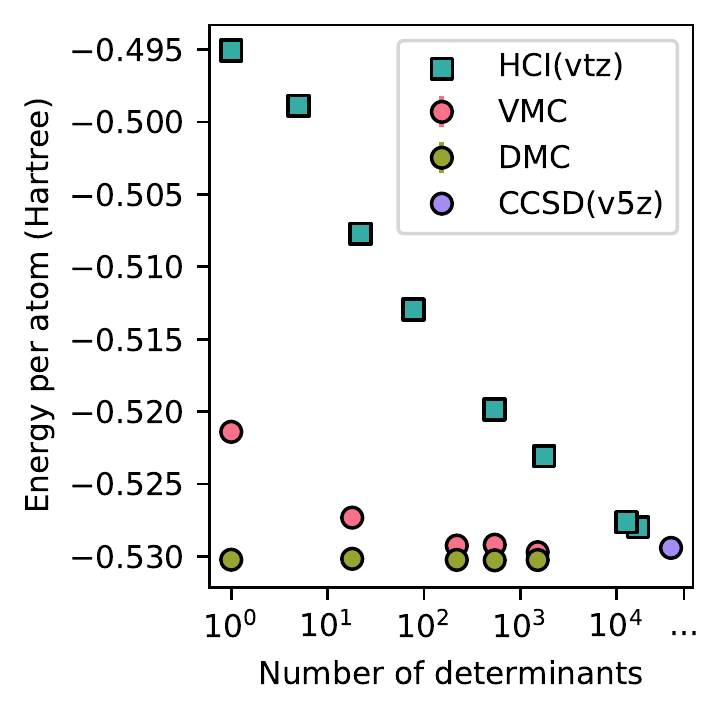}
    \caption{(H6, r=$3.0$ $a_{\text{B}}$) The energies computed using HCI wave functions (obtained using vtz basis), VMC and FN-DMC (time step = 0.02) with SJ and MSJ wave functions, and CCSD (with v5z basis) versus the number of determinants. The energies obtained agree across different methods, as the number of determinants increases.}
    \label{fig: near-exact results for a strongly correlated system}
\end{figure}

First, we check the energy convergence of our high accuracy calculations.
In Fig.~\ref{fig: near-exact results for a strongly correlated system}, we plot the ground state energies obtained using different methods (HCI, CCSD, VMC, FN-DMC) versus the number of determinants for a strongly-correlated system H6, $r=3.0$ $a_{\text{B}}$. 
Similar results are also obtained for weakly-correlated systems; data is available in the repository \cite{Workflow}.
Fig.~\ref{fig: near-exact results for a strongly correlated system} shows that the wave functions computed using FN-DMC approach yield near-exact energies with a small number of determinants. 
The energies computed using VMC, FN-DMC, and CCSD agree well as the number of determinants increases. 
The converged HCI energy is quite close to the FN-DMC and CCSD energies; however, note that we could only afford to perform converged HCI calculations at the triple-zeta level of basis.

\begin{figure*}
    \centering
    \includegraphics{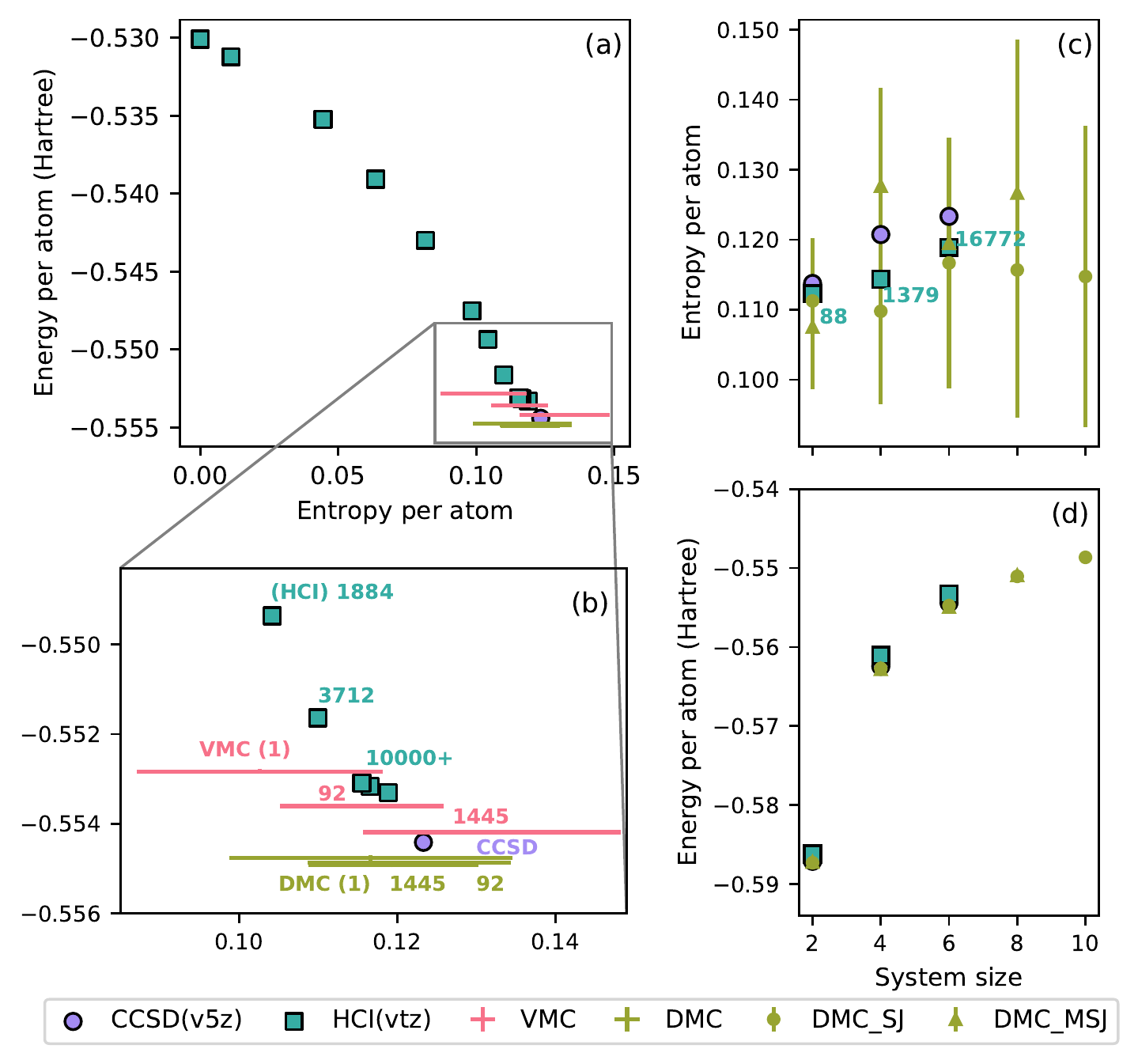}
    \caption{ (a) Converged ground state energy versus entropy per atom for a weakly-correlated system (H6, $r=1.4$ $a_{\text{B}}$) computed using CCSD (with cc-pV5Z basis), HCI wave functions (with cc-pVTZ basis), VMC and FN-DMC using SJ and MSJ trial wave functions.
    (b) Zoomed-in lower-right portion of (a). The numbers next to each point denote the numbers of determinants in the corresponding optimized wave functions. The edges of the bars on the VMC or FN-DMC points represent the lower and upper bounds computed following the method described in the method section.
    (c) The scaling of entropy per atom with system size for weakly-correlated systems ($r=1.4$ $a_{\text{B}}$). The number of determinants included in the converged HCI wave functions are annotated. 
    (d) The scaling of energy per atom with system size for weakly-correlated systems ($r=1.4$ $a_{\text{B}}$).
    }
    \label{fig: energy vs entropy for a weakly correlated system}
\end{figure*}

In Fig.~\ref{fig: energy vs entropy for a weakly correlated system}, we compare the energy and entropy of wave function methods as they converge towards the exact ground state for the weakly correlated $r=1.4$ $a_{\text{B}}$ interatomic separation. 
Like the energy, the entropy converges to a similar value for different methods. 
As one might expect for weak correlation, we find that in this case, the Jastrow factor and DMC in general are highly effective in describing the entropy of the system regardless of the number of determinants in the wave function.

\begin{figure*}
    \centering
    \includegraphics{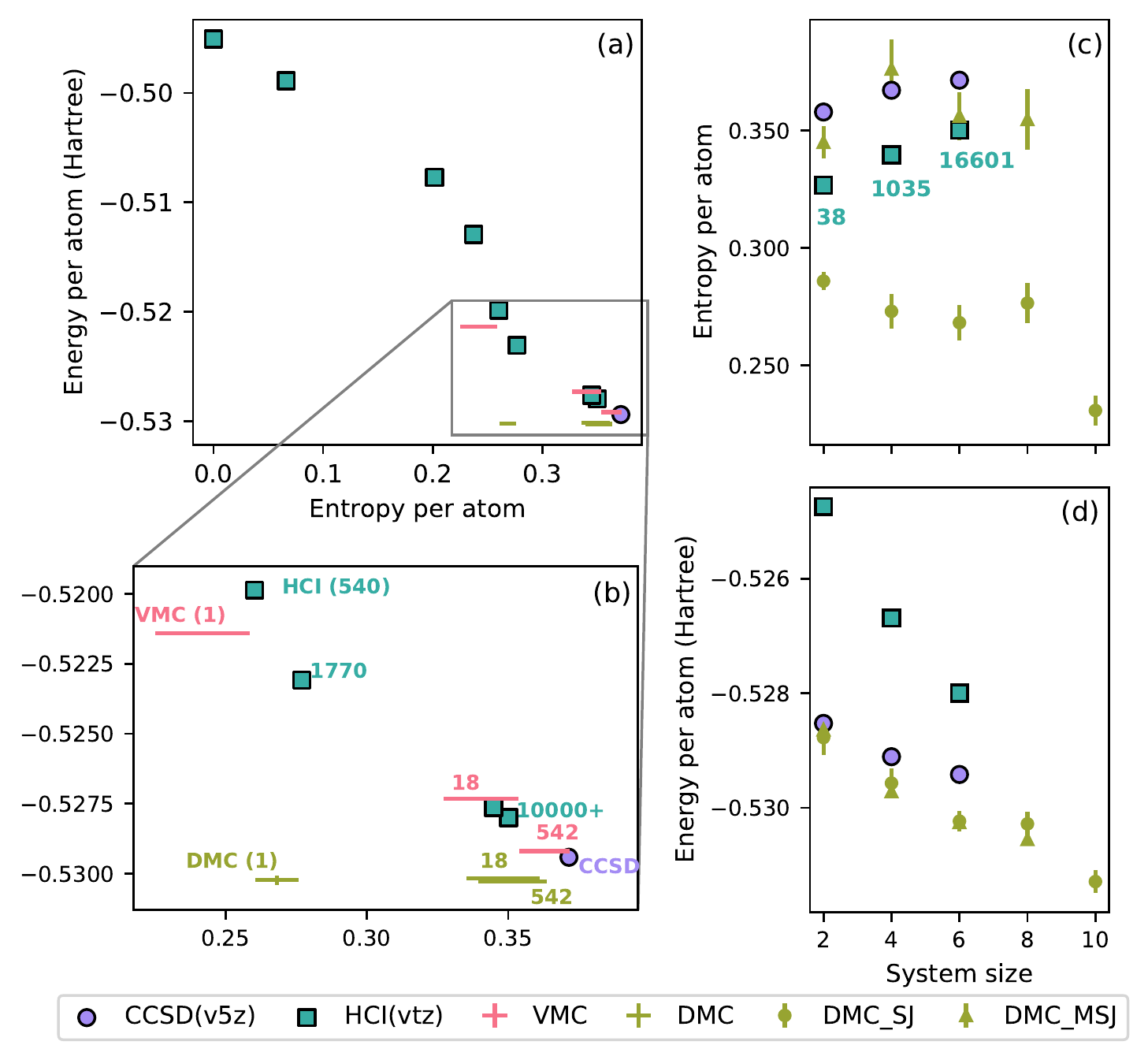}
    \caption{ (a) Converged ground state energy versus entropy per atom for a strongly-correlated system (H6, $r=3.0$ $a_{\text{B}}$) computed using CCSD (with cc-pV5Z basis), HCI wave functions (with cc-pVTZ basis), VMC and FN-DMC using SJ and MSJ trial wave functions.
    (b) Zoomed-in lower-right portion of (a).The numbers next to each point denote the numbers of determinants in the corresponding optimized wave functions. The edges of the bars on the VMC or FN-DMC points represent the lower and upper bounds computed following the method described in the method section.
    (c) The scaling of entropy per atom with system size for strongly-correlated systems ($r=3.0$ $a_{\text{B}}$). The number of determinants included in the converged HCI wave functions are annotated. 
    (d) The scaling of energy per atom with system size for strongly-correlated systems ($r=3.0$ $a_{\text{B}}$).}
    \label{fig: energy vs entropy for a strongly correlated system}
\end{figure*}

In Fig.~\ref{fig: energy vs entropy for a strongly correlated system}, we compare the energy and entropy of wave function methods as they converge towards the exact ground state for the strongly-correlated $r=3.0$ $a_{\text{B}}$ interatomic separation. 
Like weakly-correlated systems, FN-DMC with MSJ wave functions is highly effective in describing the entropy of the system regardless of the number of determinants in the wave function. 
Unlike weakly-correlated systems in which dynamic correlation is dominant, for strongly-correlated systems, FN-DMC with SJ wave functions results in energies very close to those computed from MSJ wave functions, but misses a part of the entropy which corresponds to the static correlation. 

\begin{figure}
    \centering
    \includegraphics{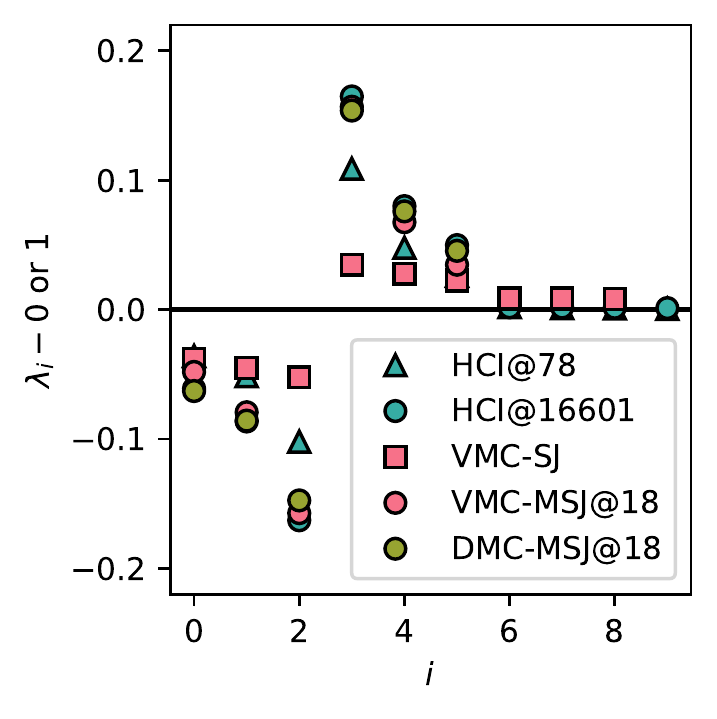}
    \caption{The entanglement spectrum of a strongly-correlated system (H6, $r=3.0$ $a_{\text{B}}$) computed using HCI wave functions with different number of determinants, VMC using optimized SJ and MSJ wave functions, and FN-DMC with MSJ trial wave function. 
    The numbers of determinants included in the wave functions are written after the names of methods. 
    The sorted eigenvalues $\lambda_i$ of the 1-RDMs (for spin up) minus either 0 or 1 are plotted against the index $i$ of the eigenvalues. Only the 10 largest eigenvalues are shown.
    }
    \label{fig: entanglement spectrum for a strongly correlated system}
\end{figure}

The difference in how the methods add correlation aligns with the ideas of dynamic and static correlation often discussed in the quantum chemistry literature.  
Dynamic correlation is identified as originating from a large set of determinants $D_{\alpha}^{\uparrow,\downarrow}$ with small coefficients $c_{\alpha}$. 
Dynamic correlation corresponds to the perturbative behavior which can be captured from a qualitatively correct, effective one-body reference state. 
Static correlation is identified from a small number of determinants $D_{\alpha}^{\uparrow,\downarrow}$  with sizable coefficients $c_{\alpha}$ towards the full many-body expansion 
\cite{MejutoZaeraStrongCorrelationThroughSCI2021}.

We further look into the entanglement spectrum (difference between eigenvalues of 1-RDM and those of idempotent matrix (1-RDM of non-interacting system)) for a strongly-correlated system (H6, $r=3.0$ $a_{\text{B}}$) shown in Fig.~\ref{fig: entanglement spectrum for a strongly correlated system}.
This figure mainly gives us 3 pieces of information.
The selected CI and Jastrow factors are complementary in their treatment of correlation. 
The selected CI wave functions first include the static correlation, then add more dynamic correlation and treat the static correlation more accurately as the expansion approaches convergence.
The Jastrow factor allows VMC and FN-DMC to treat dynamic correlation more efficiently with fewer number of determinants compared with selected CI.

Firstly, when a SJ wave function computed using VMC obtains approximately the same entropy as a HCI wave function does, they primarily treat different types of correlations. 
In Fig.~\ref{fig: entanglement spectrum for a strongly correlated system}, the entropies computed using the HCI wave function with only 78 determinants (HCI@78) and the VMC single-determinant SJ wave function (VMC-SJ) are approximately the same. 
The entanglement spectrum of HCI@78 shows more evident static correlation, as 2 of its eigenvalues of $\rho$ differ from those of an idempotent matrix.
On the other hand, VMC-SJ primarily treats dynamic correlation, as it has most of the eigenvalues closer to those of an idempotent matrix.
This information implies that the Jastrow factor reduces the wave function energy by including more dynamic correlation, which is equivalent to adding in small components of high-energy determinants into the wave function. 
Meanwhile, when reducing the wave function energy, HCI primarily treats static correlation by truncating determinants with small coefficients. 
This observation supports the idea that selected CI and Jastrow factors are complementary in their treatment of correlation \cite{dash_perturbatively_2018, dash_excited_2019, dashTailoringCIPSIExpansions2021}.

Secondly, a converged HCI wave function treats both dynamic correlation and static correlation better than an unconverged HCI wave function. 
Correspondingly, the converged HCI wave function has larger entropy than the unconverged one. 
For the system shown in Fig.~\ref{fig: entanglement spectrum for a strongly correlated system}, HCI wave functions converges when using 16601 determinants (HCI@16601). 
Compared with the $\rho$ of HCI@78 that has only a few non-zero eigenvalues, the $\rho$ of HCI@16601 has many small eigenvalues which are not plotted here. 
In addition, the first few eigenvalues of HCI@78 are closer to those of an idempotent matrix than those of HCI@16601. 
This information implies that when CI wave functions approach to convergence, they add more dynamic correlation and also treat static correlation more accurately.

Thirdly, FN-DMC treats the correlation efficiently using a trial wave function with only a few optimized determinants from a converged HCI wave function. 
As shown in Fig.~\ref{fig: entanglement spectrum for a strongly correlated system}, the entanglement spectra of DMC-MSJ@18 and HCI@16601 are very similar. 
This DMC-MSJ wave function optimizes only the largest 18 of 16601 determinants computed by HCI wave function, but DMC-MSJ obtains almost the same entropy as the HCI wave function.


\section*{Conclusion} 

In conclusion, we used multipartite entanglement and its spectrum to evaluate the differences between quantum chemistry and quantum Monte Carlo approaches to electron correlation. 
We developed a new algorithm to evaluate the entropy of randomized matrices, the circle reject method, which enabled an accurate evaluation of this quantity using quantum Monte Carlo. 
We found that the Jastrow factor indeed appears to mainly add dynamic correlation by creating many small eigenvalues of the 1-RDM, while selected CI methods tend to create a few large eigenvalues first, which is an explicit observation of the complementary nature of these terms in the wave function. 

The circle reject algorithm could find more uses, as the use of stochastic algorithms in quantum chemistry appears to be increasing \cite{morales-silvaFrontiersStochasticElectronic2021}.
It is particularly worth using if a matrix is likely to have very few non-zero eigenvalues, but is evaluated stochastically.

The eigenvalues of the 1-RDM are simple to compute, and we believe that it should become more standard to evaluate the multipartite entanglement as one measure of electronic correlation. 
Such a measure also allows one to make contact with the roughly equivalent homogeneous electron gas, since the momentum distribution is known for several values of $r_s$ \cite{holzmannMomentumDistributionHomogeneous2011b}.

\section*{Acknowledgements}
L.K.W. was supported by U.S. National Science Foundation via Award No. 1931258 to implement methodology in the \texttt{PyQMC} project. 
Y.C. was supported by the U.S. Department of Energy, Office of Science, Office of Basic Energy Sciences, Computational Materials Sciences Program, under Award No. DE-SC0020177 to assist with analysis and writing of the manuscript.
This work made use of the Illinois Campus Cluster, a computing resource that is operated by the Illinois Campus Cluster Program (ICCP) in conjunction with the National Center for Supercomputing Applications (NCSA) and which is supported by funds from the University of Illinois at Urbana-Champaign.

\bibliography{ref.bib}

\begin{thebibliography}{40}%
\makeatletter
\providecommand \@ifxundefined [1]{%
 \@ifx{#1\undefined}
}%
\providecommand \@ifnum [1]{%
 \ifnum #1\expandafter \@firstoftwo
 \else \expandafter \@secondoftwo
 \fi
}%
\providecommand \@ifx [1]{%
 \ifx #1\expandafter \@firstoftwo
 \else \expandafter \@secondoftwo
 \fi
}%
\providecommand \natexlab [1]{#1}%
\providecommand \enquote  [1]{``#1''}%
\providecommand \bibnamefont  [1]{#1}%
\providecommand \bibfnamefont [1]{#1}%
\providecommand \citenamefont [1]{#1}%
\providecommand \href@noop [0]{\@secondoftwo}%
\providecommand \href [0]{\begingroup \@sanitize@url \@href}%
\providecommand \@href[1]{\@@startlink{#1}\@@href}%
\providecommand \@@href[1]{\endgroup#1\@@endlink}%
\providecommand \@sanitize@url [0]{\catcode `\\12\catcode `\$12\catcode
  `\&12\catcode `\#12\catcode `\^12\catcode `\_12\catcode `\%12\relax}%
\providecommand \@@startlink[1]{}%
\providecommand \@@endlink[0]{}%
\providecommand \url  [0]{\begingroup\@sanitize@url \@url }%
\providecommand \@url [1]{\endgroup\@href {#1}{\urlprefix }}%
\providecommand \urlprefix  [0]{URL }%
\providecommand \Eprint [0]{\href }%
\providecommand \doibase [0]{http://dx.doi.org/}%
\providecommand \selectlanguage [0]{\@gobble}%
\providecommand \bibinfo  [0]{\@secondoftwo}%
\providecommand \bibfield  [0]{\@secondoftwo}%
\providecommand \translation [1]{[#1]}%
\providecommand \BibitemOpen [0]{}%
\providecommand \bibitemStop [0]{}%
\providecommand \bibitemNoStop [0]{.\EOS\space}%
\providecommand \EOS [0]{\spacefactor3000\relax}%
\providecommand \BibitemShut  [1]{\csname bibitem#1\endcsname}%
\let\auto@bib@innerbib\@empty
\bibitem [{\citenamefont {Sun}\ \emph {et~al.}(2015)\citenamefont {Sun},
  \citenamefont {Ruzsinszky},\ and\ \citenamefont
  {Perdew}}]{sun_strongly_2015}%
  \BibitemOpen
  \bibfield  {author} {\bibinfo {author} {\bibfnamefont {J.}~\bibnamefont
  {Sun}}, \bibinfo {author} {\bibfnamefont {A.}~\bibnamefont {Ruzsinszky}}, \
  and\ \bibinfo {author} {\bibfnamefont {J.}~\bibnamefont {Perdew}},\ }\href
  {\doibase 10.1103/PhysRevLett.115.036402} {\bibfield  {journal} {\bibinfo
  {journal} {Physical Review Letters}\ }\textbf {\bibinfo {volume} {115}},\
  \bibinfo {pages} {036402} (\bibinfo {year} {2015})}\BibitemShut {NoStop}%
\bibitem [{\citenamefont {Sun}\ \emph {et~al.}(2016)\citenamefont {Sun},
  \citenamefont {Remsing}, \citenamefont {Zhang}, \citenamefont {Sun},
  \citenamefont {Ruzsinszky}, \citenamefont {Peng}, \citenamefont {Yang},
  \citenamefont {Paul}, \citenamefont {Waghmare}, \citenamefont {Wu},
  \citenamefont {Klein},\ and\ \citenamefont {Perdew}}]{sun_accurate_2016}%
  \BibitemOpen
  \bibfield  {author} {\bibinfo {author} {\bibfnamefont {J.}~\bibnamefont
  {Sun}}, \bibinfo {author} {\bibfnamefont {R.~C.}\ \bibnamefont {Remsing}},
  \bibinfo {author} {\bibfnamefont {Y.}~\bibnamefont {Zhang}}, \bibinfo
  {author} {\bibfnamefont {Z.}~\bibnamefont {Sun}}, \bibinfo {author}
  {\bibfnamefont {A.}~\bibnamefont {Ruzsinszky}}, \bibinfo {author}
  {\bibfnamefont {H.}~\bibnamefont {Peng}}, \bibinfo {author} {\bibfnamefont
  {Z.}~\bibnamefont {Yang}}, \bibinfo {author} {\bibfnamefont {A.}~\bibnamefont
  {Paul}}, \bibinfo {author} {\bibfnamefont {U.}~\bibnamefont {Waghmare}},
  \bibinfo {author} {\bibfnamefont {X.}~\bibnamefont {Wu}}, \bibinfo {author}
  {\bibfnamefont {M.~L.}\ \bibnamefont {Klein}}, \ and\ \bibinfo {author}
  {\bibfnamefont {J.~P.}\ \bibnamefont {Perdew}},\ }\href {\doibase
  10.1038/nchem.2535} {\bibfield  {journal} {\bibinfo  {journal} {Nature
  Chemistry}\ }\textbf {\bibinfo {volume} {8}},\ \bibinfo {pages} {831}
  (\bibinfo {year} {2016})}\BibitemShut {NoStop}%
\bibitem [{\citenamefont {Morales}\ \emph {et~al.}(2014)\citenamefont
  {Morales}, \citenamefont {Gergely}, \citenamefont {McMinis}, \citenamefont
  {McMahon}, \citenamefont {Kim},\ and\ \citenamefont
  {Ceperley}}]{morales_quantum_2014}%
  \BibitemOpen
  \bibfield  {author} {\bibinfo {author} {\bibfnamefont {M.~A.}\ \bibnamefont
  {Morales}}, \bibinfo {author} {\bibfnamefont {J.~R.}\ \bibnamefont
  {Gergely}}, \bibinfo {author} {\bibfnamefont {J.}~\bibnamefont {McMinis}},
  \bibinfo {author} {\bibfnamefont {J.~M.}\ \bibnamefont {McMahon}}, \bibinfo
  {author} {\bibfnamefont {J.}~\bibnamefont {Kim}}, \ and\ \bibinfo {author}
  {\bibfnamefont {D.~M.}\ \bibnamefont {Ceperley}},\ }\href {\doibase
  10.1021/ct500129p} {\bibfield  {journal} {\bibinfo  {journal} {Journal of
  Chemical Theory and Computation}\ }\textbf {\bibinfo {volume} {10}},\
  \bibinfo {pages} {2355} (\bibinfo {year} {2014})}\BibitemShut {NoStop}%
\bibitem [{\citenamefont {Borlido}\ \emph {et~al.}(2019)\citenamefont
  {Borlido}, \citenamefont {Aull}, \citenamefont {Huran}, \citenamefont {Tran},
  \citenamefont {Marques},\ and\ \citenamefont
  {Botti}}]{borlido_large-scale_2019}%
  \BibitemOpen
  \bibfield  {author} {\bibinfo {author} {\bibfnamefont {P.}~\bibnamefont
  {Borlido}}, \bibinfo {author} {\bibfnamefont {T.}~\bibnamefont {Aull}},
  \bibinfo {author} {\bibfnamefont {A.~W.}\ \bibnamefont {Huran}}, \bibinfo
  {author} {\bibfnamefont {F.}~\bibnamefont {Tran}}, \bibinfo {author}
  {\bibfnamefont {M.~A.~L.}\ \bibnamefont {Marques}}, \ and\ \bibinfo {author}
  {\bibfnamefont {S.}~\bibnamefont {Botti}},\ }\href {\doibase
  10.1021/acs.jctc.9b00322} {\bibfield  {journal} {\bibinfo  {journal} {Journal
  of Chemical Theory and Computation}\ }\textbf {\bibinfo {volume} {15}},\
  \bibinfo {pages} {5069} (\bibinfo {year} {2019})}\BibitemShut {NoStop}%
\bibitem [{\citenamefont {Borlido}\ \emph {et~al.}(2020)\citenamefont
  {Borlido}, \citenamefont {Schmidt}, \citenamefont {Huran}, \citenamefont
  {Tran}, \citenamefont {Marques},\ and\ \citenamefont
  {Botti}}]{borlido_exchange-correlation_2020}%
  \BibitemOpen
  \bibfield  {author} {\bibinfo {author} {\bibfnamefont {P.}~\bibnamefont
  {Borlido}}, \bibinfo {author} {\bibfnamefont {J.}~\bibnamefont {Schmidt}},
  \bibinfo {author} {\bibfnamefont {A.~W.}\ \bibnamefont {Huran}}, \bibinfo
  {author} {\bibfnamefont {F.}~\bibnamefont {Tran}}, \bibinfo {author}
  {\bibfnamefont {M.~A.~L.}\ \bibnamefont {Marques}}, \ and\ \bibinfo {author}
  {\bibfnamefont {S.}~\bibnamefont {Botti}},\ }\href {\doibase
  10.1038/s41524-020-00360-0} {\bibfield  {journal} {\bibinfo  {journal} {npj
  Computational Materials}\ }\textbf {\bibinfo {volume} {6}},\ \bibinfo {pages}
  {96} (\bibinfo {year} {2020})}\BibitemShut {NoStop}%
\bibitem [{\citenamefont {Ceperley}\ and\ \citenamefont
  {Alder}(1980)}]{ceperley_ground_1980}%
  \BibitemOpen
  \bibfield  {author} {\bibinfo {author} {\bibfnamefont {D.~M.}\ \bibnamefont
  {Ceperley}}\ and\ \bibinfo {author} {\bibfnamefont {B.~J.}\ \bibnamefont
  {Alder}},\ }\href {\doibase 10.1103/PhysRevLett.45.566} {\bibfield  {journal}
  {\bibinfo  {journal} {Physical Review Letters}\ }\textbf {\bibinfo {volume}
  {45}},\ \bibinfo {pages} {566} (\bibinfo {year} {1980})}\BibitemShut
  {NoStop}%
\bibitem [{\citenamefont {Foulkes}\ \emph {et~al.}(2001)\citenamefont
  {Foulkes}, \citenamefont {Mitas}, \citenamefont {Needs},\ and\ \citenamefont
  {Rajagopal}}]{Foulkes}%
  \BibitemOpen
  \bibfield  {author} {\bibinfo {author} {\bibfnamefont {W.~M.~C.}\
  \bibnamefont {Foulkes}}, \bibinfo {author} {\bibfnamefont {L.}~\bibnamefont
  {Mitas}}, \bibinfo {author} {\bibfnamefont {R.~J.}\ \bibnamefont {Needs}}, \
  and\ \bibinfo {author} {\bibfnamefont {G.}~\bibnamefont {Rajagopal}},\ }\href
  {\doibase 10.1103/RevModPhys.73.33} {\bibfield  {journal} {\bibinfo
  {journal} {Reviews of Modern Physics}\ }\textbf {\bibinfo {volume} {73}},\
  \bibinfo {pages} {33} (\bibinfo {year} {2001})}\BibitemShut {NoStop}%
\bibitem [{\citenamefont {Cizek}\ and\ \citenamefont
  {Paldus}(1980)}]{cizek_coupled_1980}%
  \BibitemOpen
  \bibfield  {author} {\bibinfo {author} {\bibfnamefont {J.}~\bibnamefont
  {Cizek}}\ and\ \bibinfo {author} {\bibfnamefont {J.}~\bibnamefont {Paldus}},\
  }\href {\doibase 10.1088/0031-8949/21/3-4/006} {\bibfield  {journal}
  {\bibinfo  {journal} {Physica Scripta}\ }\textbf {\bibinfo {volume} {21}},\
  \bibinfo {pages} {251} (\bibinfo {year} {1980})}\BibitemShut {NoStop}%
\bibitem [{\citenamefont {Paldus}\ \emph {et~al.}(1989)\citenamefont {Paldus},
  \citenamefont {Čížek},\ and\ \citenamefont
  {Jeziorski}}]{paldus_coupled_1989}%
  \BibitemOpen
  \bibfield  {author} {\bibinfo {author} {\bibfnamefont {J.}~\bibnamefont
  {Paldus}}, \bibinfo {author} {\bibfnamefont {J.}~\bibnamefont {Čížek}}, \
  and\ \bibinfo {author} {\bibfnamefont {B.}~\bibnamefont {Jeziorski}},\ }\href
  {\doibase 10.1063/1.456647} {\bibfield  {journal} {\bibinfo  {journal} {The
  Journal of Chemical Physics}\ }\textbf {\bibinfo {volume} {90}},\ \bibinfo
  {pages} {4356} (\bibinfo {year} {1989})}\BibitemShut {NoStop}%
\bibitem [{\citenamefont {Bartlett}\ and\ \citenamefont
  {Musiał}(2007)}]{bartlett_coupled-cluster_2007}%
  \BibitemOpen
  \bibfield  {author} {\bibinfo {author} {\bibfnamefont {R.~J.}\ \bibnamefont
  {Bartlett}}\ and\ \bibinfo {author} {\bibfnamefont {M.}~\bibnamefont
  {Musiał}},\ }\href {\doibase 10.1103/RevModPhys.79.291} {\bibfield
  {journal} {\bibinfo  {journal} {Reviews of Modern Physics}\ }\textbf
  {\bibinfo {volume} {79}},\ \bibinfo {pages} {291} (\bibinfo {year}
  {2007})}\BibitemShut {NoStop}%
\bibitem [{\citenamefont {White}(1992)}]{whiteDensityMatrixFormulation1992}%
  \BibitemOpen
  \bibfield  {author} {\bibinfo {author} {\bibfnamefont {S.~R.}\ \bibnamefont
  {White}},\ }\href {\doibase 10.1103/PhysRevLett.69.2863} {\bibfield
  {journal} {\bibinfo  {journal} {Physical Review Letters}\ }\textbf {\bibinfo
  {volume} {69}},\ \bibinfo {pages} {2863} (\bibinfo {year}
  {1992})}\BibitemShut {NoStop}%
\bibitem [{\citenamefont {Schollwöck}(2005)}]{schollwock_density-matrix_2005}%
  \BibitemOpen
  \bibfield  {author} {\bibinfo {author} {\bibfnamefont {U.}~\bibnamefont
  {Schollwöck}},\ }\href
  {https://journals.aps.org/rmp/abstract/10.1103/RevModPhys.77.259} {\bibfield
  {journal} {\bibinfo  {journal} {Reviews of Modern Physics}\ }\textbf
  {\bibinfo {volume} {77}},\ \bibinfo {pages} {57} (\bibinfo {year}
  {2005})}\BibitemShut {NoStop}%
\bibitem [{\citenamefont {Huron}\ \emph {et~al.}(1973)\citenamefont {Huron},
  \citenamefont {Malrieu},\ and\ \citenamefont
  {Rancurel}}]{huron_iterative_1973}%
  \BibitemOpen
  \bibfield  {author} {\bibinfo {author} {\bibfnamefont {B.}~\bibnamefont
  {Huron}}, \bibinfo {author} {\bibfnamefont {J.~P.}\ \bibnamefont {Malrieu}},
  \ and\ \bibinfo {author} {\bibfnamefont {P.}~\bibnamefont {Rancurel}},\
  }\href {\doibase 10.1063/1.1679199} {\bibfield  {journal} {\bibinfo
  {journal} {The Journal of Chemical Physics}\ }\textbf {\bibinfo {volume}
  {58}},\ \bibinfo {pages} {5745} (\bibinfo {year} {1973})}\BibitemShut
  {NoStop}%
\bibitem [{\citenamefont {Cimiraglia}\ and\ \citenamefont
  {Persico}(1987)}]{cimiraglia_recent_1987}%
  \BibitemOpen
  \bibfield  {author} {\bibinfo {author} {\bibfnamefont {R.}~\bibnamefont
  {Cimiraglia}}\ and\ \bibinfo {author} {\bibfnamefont {M.}~\bibnamefont
  {Persico}},\ }\href {\doibase 10.1002/jcc.540080105} {\bibfield  {journal}
  {\bibinfo  {journal} {Journal of Computational Chemistry}\ }\textbf {\bibinfo
  {volume} {8}},\ \bibinfo {pages} {39} (\bibinfo {year} {1987})}\BibitemShut
  {NoStop}%
\bibitem [{\citenamefont {Holmes}\ \emph {et~al.}(2016)\citenamefont {Holmes},
  \citenamefont {Tubman},\ and\ \citenamefont {Umrigar}}]{HCI}%
  \BibitemOpen
  \bibfield  {author} {\bibinfo {author} {\bibfnamefont {A.~A.}\ \bibnamefont
  {Holmes}}, \bibinfo {author} {\bibfnamefont {N.~M.}\ \bibnamefont {Tubman}},
  \ and\ \bibinfo {author} {\bibfnamefont {C.~J.}\ \bibnamefont {Umrigar}},\
  }\href {\doibase 10.1021/acs.jctc.6b00407} {\bibfield  {journal} {\bibinfo
  {journal} {Journal of Chemical Theory and Computation}\ }\textbf {\bibinfo
  {volume} {12}},\ \bibinfo {pages} {3674} (\bibinfo {year}
  {2016})}\BibitemShut {NoStop}%
\bibitem [{\citenamefont {Wolpert}\ and\ \citenamefont
  {Macready}(1997)}]{wolpertNoFreeLunch1997}%
  \BibitemOpen
  \bibfield  {author} {\bibinfo {author} {\bibfnamefont {D.}~\bibnamefont
  {Wolpert}}\ and\ \bibinfo {author} {\bibfnamefont {W.}~\bibnamefont
  {Macready}},\ }\href {\doibase 10.1109/4235.585893} {\bibfield  {journal}
  {\bibinfo  {journal} {IEEE Transactions on Evolutionary Computation}\
  }\textbf {\bibinfo {volume} {1}},\ \bibinfo {pages} {67} (\bibinfo {year}
  {1997})}\BibitemShut {NoStop}%
\bibitem [{\citenamefont {Troyer}\ and\ \citenamefont
  {Wiese}(2005)}]{troyerComputationalComplexityFundamental2005}%
  \BibitemOpen
  \bibfield  {author} {\bibinfo {author} {\bibfnamefont {M.}~\bibnamefont
  {Troyer}}\ and\ \bibinfo {author} {\bibfnamefont {U.-J.}\ \bibnamefont
  {Wiese}},\ }\href {\doibase 10.1103/PhysRevLett.94.170201} {\bibfield
  {journal} {\bibinfo  {journal} {Physical Review Letters}\ }\textbf {\bibinfo
  {volume} {94}},\ \bibinfo {pages} {170201} (\bibinfo {year}
  {2005})}\BibitemShut {NoStop}%
\bibitem [{\citenamefont {{Simons Collaboration on the Many-Electron Problem}}\
  \emph {et~al.}(2020)\citenamefont {{Simons Collaboration on the Many-Electron
  Problem}}, \citenamefont {Williams}, \citenamefont {Yao}, \citenamefont {Li},
  \citenamefont {Chen}, \citenamefont {Shi}, \citenamefont {Motta},
  \citenamefont {Niu}, \citenamefont {Ray}, \citenamefont {Guo}, \citenamefont
  {Anderson}, \citenamefont {Li}, \citenamefont {Tran}, \citenamefont {Yeh},
  \citenamefont {Mussard}, \citenamefont {Sharma}, \citenamefont {Bruneval},
  \citenamefont {{van Schilfgaarde}}, \citenamefont {Booth}, \citenamefont
  {Chan}, \citenamefont {Zhang}, \citenamefont {Gull}, \citenamefont {Zgid},
  \citenamefont {Millis}, \citenamefont {Umrigar},\ and\ \citenamefont
  {Wagner}}]{simonscollaborationonthemany-electronproblemDirectComparisonManyBody2020}%
  \BibitemOpen
  \bibfield  {author} {\bibinfo {author} {\bibnamefont {{Simons Collaboration
  on the Many-Electron Problem}}}, \bibinfo {author} {\bibfnamefont {K.~T.}\
  \bibnamefont {Williams}}, \bibinfo {author} {\bibfnamefont {Y.}~\bibnamefont
  {Yao}}, \bibinfo {author} {\bibfnamefont {J.}~\bibnamefont {Li}}, \bibinfo
  {author} {\bibfnamefont {L.}~\bibnamefont {Chen}}, \bibinfo {author}
  {\bibfnamefont {H.}~\bibnamefont {Shi}}, \bibinfo {author} {\bibfnamefont
  {M.}~\bibnamefont {Motta}}, \bibinfo {author} {\bibfnamefont
  {C.}~\bibnamefont {Niu}}, \bibinfo {author} {\bibfnamefont {U.}~\bibnamefont
  {Ray}}, \bibinfo {author} {\bibfnamefont {S.}~\bibnamefont {Guo}}, \bibinfo
  {author} {\bibfnamefont {R.~J.}\ \bibnamefont {Anderson}}, \bibinfo {author}
  {\bibfnamefont {J.}~\bibnamefont {Li}}, \bibinfo {author} {\bibfnamefont
  {L.~N.}\ \bibnamefont {Tran}}, \bibinfo {author} {\bibfnamefont {C.-N.}\
  \bibnamefont {Yeh}}, \bibinfo {author} {\bibfnamefont {B.}~\bibnamefont
  {Mussard}}, \bibinfo {author} {\bibfnamefont {S.}~\bibnamefont {Sharma}},
  \bibinfo {author} {\bibfnamefont {F.}~\bibnamefont {Bruneval}}, \bibinfo
  {author} {\bibfnamefont {M.}~\bibnamefont {{van Schilfgaarde}}}, \bibinfo
  {author} {\bibfnamefont {G.~H.}\ \bibnamefont {Booth}}, \bibinfo {author}
  {\bibfnamefont {G.~K.-L.}\ \bibnamefont {Chan}}, \bibinfo {author}
  {\bibfnamefont {S.}~\bibnamefont {Zhang}}, \bibinfo {author} {\bibfnamefont
  {E.}~\bibnamefont {Gull}}, \bibinfo {author} {\bibfnamefont {D.}~\bibnamefont
  {Zgid}}, \bibinfo {author} {\bibfnamefont {A.}~\bibnamefont {Millis}},
  \bibinfo {author} {\bibfnamefont {C.~J.}\ \bibnamefont {Umrigar}}, \ and\
  \bibinfo {author} {\bibfnamefont {L.~K.}\ \bibnamefont {Wagner}},\ }\href
  {\doibase 10.1103/PhysRevX.10.011041} {\bibfield  {journal} {\bibinfo
  {journal} {Physical Review X}\ }\textbf {\bibinfo {volume} {10}},\ \bibinfo
  {pages} {011041} (\bibinfo {year} {2020})}\BibitemShut {NoStop}%
\bibitem [{\citenamefont {Amico}\ \emph {et~al.}(2008)\citenamefont {Amico},
  \citenamefont {Fazio}, \citenamefont {Osterloh},\ and\ \citenamefont
  {Vedral}}]{amico_entanglement_2008}%
  \BibitemOpen
  \bibfield  {author} {\bibinfo {author} {\bibfnamefont {L.}~\bibnamefont
  {Amico}}, \bibinfo {author} {\bibfnamefont {R.}~\bibnamefont {Fazio}},
  \bibinfo {author} {\bibfnamefont {A.}~\bibnamefont {Osterloh}}, \ and\
  \bibinfo {author} {\bibfnamefont {V.}~\bibnamefont {Vedral}},\ }\href
  {\doibase 10.1103/RevModPhys.80.517} {\bibfield  {journal} {\bibinfo
  {journal} {Reviews of Modern Physics}\ }\textbf {\bibinfo {volume} {80}},\
  \bibinfo {pages} {517} (\bibinfo {year} {2008})}\BibitemShut {NoStop}%
\bibitem [{\citenamefont {Vidal}(2003)}]{vidal_efficient_2003}%
  \BibitemOpen
  \bibfield  {author} {\bibinfo {author} {\bibfnamefont {G.}~\bibnamefont
  {Vidal}},\ }\href {\doibase 10.1103/PhysRevLett.91.147902} {\bibfield
  {journal} {\bibinfo  {journal} {Physical Review Letters}\ }\textbf {\bibinfo
  {volume} {91}},\ \bibinfo {pages} {147902} (\bibinfo {year}
  {2003})}\BibitemShut {NoStop}%
\bibitem [{\citenamefont {Vidal}(2004)}]{vidal_efficient_2004}%
  \BibitemOpen
  \bibfield  {author} {\bibinfo {author} {\bibfnamefont {G.}~\bibnamefont
  {Vidal}},\ }\href {\doibase 10.1103/PhysRevLett.93.040502} {\bibfield
  {journal} {\bibinfo  {journal} {Physical Review Letters}\ }\textbf {\bibinfo
  {volume} {93}},\ \bibinfo {pages} {040502} (\bibinfo {year}
  {2004})}\BibitemShut {NoStop}%
\bibitem [{\citenamefont {McMinis}\ and\ \citenamefont
  {Tubman}(2013)}]{mcminisRenyiEntropyInteracting2013}%
  \BibitemOpen
  \bibfield  {author} {\bibinfo {author} {\bibfnamefont {J.}~\bibnamefont
  {McMinis}}\ and\ \bibinfo {author} {\bibfnamefont {N.~M.}\ \bibnamefont
  {Tubman}},\ }\href {\doibase 10.1103/PhysRevB.87.081108} {\bibfield
  {journal} {\bibinfo  {journal} {Physical Review B}\ }\textbf {\bibinfo
  {volume} {87}},\ \bibinfo {pages} {081108} (\bibinfo {year}
  {2013})}\BibitemShut {NoStop}%
\bibitem [{\citenamefont {Juh{\'a}sz}\ and\ \citenamefont
  {Mazziotti}(2006)}]{juhaszCumulantTwoparticleReduced2006b}%
  \BibitemOpen
  \bibfield  {author} {\bibinfo {author} {\bibfnamefont {T.}~\bibnamefont
  {Juh{\'a}sz}}\ and\ \bibinfo {author} {\bibfnamefont {D.~A.}\ \bibnamefont
  {Mazziotti}},\ }\href {\doibase 10.1063/1.2378768} {\bibfield  {journal}
  {\bibinfo  {journal} {The Journal of Chemical Physics}\ }\textbf {\bibinfo
  {volume} {125}},\ \bibinfo {pages} {174105} (\bibinfo {year}
  {2006})}\BibitemShut {NoStop}%
\bibitem [{\citenamefont
  {Delle~Site}(2015)}]{dellesiteShannonEntropyManyelectron2015a}%
  \BibitemOpen
  \bibfield  {author} {\bibinfo {author} {\bibfnamefont {L.}~\bibnamefont
  {Delle~Site}},\ }\href {\doibase 10.1002/qua.24823} {\bibfield  {journal}
  {\bibinfo  {journal} {International Journal of Quantum Chemistry}\ }\textbf
  {\bibinfo {volume} {115}},\ \bibinfo {pages} {1396} (\bibinfo {year}
  {2015})}\BibitemShut {NoStop}%
\bibitem [{\citenamefont {{Ramos-Cordoba}}\ \emph {et~al.}(2016)\citenamefont
  {{Ramos-Cordoba}}, \citenamefont {Salvador},\ and\ \citenamefont
  {Matito}}]{ramos-cordobaSeparationDynamicNondynamic2016}%
  \BibitemOpen
  \bibfield  {author} {\bibinfo {author} {\bibfnamefont {E.}~\bibnamefont
  {{Ramos-Cordoba}}}, \bibinfo {author} {\bibfnamefont {P.}~\bibnamefont
  {Salvador}}, \ and\ \bibinfo {author} {\bibfnamefont {E.}~\bibnamefont
  {Matito}},\ }\href {\doibase 10.1039/C6CP03072F} {\bibfield  {journal}
  {\bibinfo  {journal} {Physical Chemistry Chemical Physics}\ }\textbf
  {\bibinfo {volume} {18}},\ \bibinfo {pages} {24015} (\bibinfo {year}
  {2016})}\BibitemShut {NoStop}%
\bibitem [{\citenamefont {Boguslawski}\ \emph {et~al.}(2012)\citenamefont
  {Boguslawski}, \citenamefont {Tecmer}, \citenamefont {Legeza},\ and\
  \citenamefont {Reiher}}]{boguslawskiEntanglementMeasuresSingle2012}%
  \BibitemOpen
  \bibfield  {author} {\bibinfo {author} {\bibfnamefont {K.}~\bibnamefont
  {Boguslawski}}, \bibinfo {author} {\bibfnamefont {P.}~\bibnamefont {Tecmer}},
  \bibinfo {author} {\bibfnamefont {{\"O}.}~\bibnamefont {Legeza}}, \ and\
  \bibinfo {author} {\bibfnamefont {M.}~\bibnamefont {Reiher}},\ }\href
  {\doibase 10.1021/jz301319v} {\bibfield  {journal} {\bibinfo  {journal} {The
  Journal of Physical Chemistry Letters}\ }\textbf {\bibinfo {volume} {3}},\
  \bibinfo {pages} {3129} (\bibinfo {year} {2012})}\BibitemShut {NoStop}%
\bibitem [{\citenamefont {Filippi}\ and\ \citenamefont
  {Umrigar}(1996)}]{filippiMulticonfigurationWaveFunctions1996}%
  \BibitemOpen
  \bibfield  {author} {\bibinfo {author} {\bibfnamefont {C.}~\bibnamefont
  {Filippi}}\ and\ \bibinfo {author} {\bibfnamefont {C.~J.}\ \bibnamefont
  {Umrigar}},\ }\href {\doibase 10.1063/1.471865} {\bibfield  {journal}
  {\bibinfo  {journal} {The Journal of Chemical Physics}\ }\textbf {\bibinfo
  {volume} {105}},\ \bibinfo {pages} {213} (\bibinfo {year}
  {1996})}\BibitemShut {NoStop}%
\bibitem [{\citenamefont {Holzmann}\ \emph {et~al.}(2011)\citenamefont
  {Holzmann}, \citenamefont {Bernu}, \citenamefont {Pierleoni}, \citenamefont
  {McMinis}, \citenamefont {Ceperley}, \citenamefont {Olevano},\ and\
  \citenamefont {Delle~Site}}]{holzmannMomentumDistributionHomogeneous2011b}%
  \BibitemOpen
  \bibfield  {author} {\bibinfo {author} {\bibfnamefont {M.}~\bibnamefont
  {Holzmann}}, \bibinfo {author} {\bibfnamefont {B.}~\bibnamefont {Bernu}},
  \bibinfo {author} {\bibfnamefont {C.}~\bibnamefont {Pierleoni}}, \bibinfo
  {author} {\bibfnamefont {J.}~\bibnamefont {McMinis}}, \bibinfo {author}
  {\bibfnamefont {D.~M.}\ \bibnamefont {Ceperley}}, \bibinfo {author}
  {\bibfnamefont {V.}~\bibnamefont {Olevano}}, \ and\ \bibinfo {author}
  {\bibfnamefont {L.}~\bibnamefont {Delle~Site}},\ }\href {\doibase
  10.1103/PhysRevLett.107.110402} {\bibfield  {journal} {\bibinfo  {journal}
  {Physical Review Letters}\ }\textbf {\bibinfo {volume} {107}},\ \bibinfo
  {pages} {110402} (\bibinfo {year} {2011})}\BibitemShut {NoStop}%
\bibitem [{\citenamefont {Dunning}(1989)}]{Basis}%
  \BibitemOpen
  \bibfield  {author} {\bibinfo {author} {\bibfnamefont {T.~H.}\ \bibnamefont
  {Dunning}},\ }\href {\doibase 10.1063/1.456153} {\bibfield  {journal}
  {\bibinfo  {journal} {The Journal of Chemical Physics}\ }\textbf {\bibinfo
  {volume} {90}},\ \bibinfo {pages} {1007} (\bibinfo {year}
  {1989})}\BibitemShut {NoStop}%
\bibitem [{\citenamefont {Wagner}\ \emph {et~al.}(2009)\citenamefont {Wagner},
  \citenamefont {Bajdich},\ and\ \citenamefont {Mitas}}]{Jastrow}%
  \BibitemOpen
  \bibfield  {author} {\bibinfo {author} {\bibfnamefont {L.~K.}\ \bibnamefont
  {Wagner}}, \bibinfo {author} {\bibfnamefont {M.}~\bibnamefont {Bajdich}}, \
  and\ \bibinfo {author} {\bibfnamefont {L.}~\bibnamefont {Mitas}},\ }\href
  {\doibase 10.1016/j.jcp.2009.01.017} {\bibfield  {journal} {\bibinfo
  {journal} {Journal of Computational Physics}\ }\textbf {\bibinfo {volume}
  {228}},\ \bibinfo {pages} {3390} (\bibinfo {year} {2009})}\BibitemShut
  {NoStop}%
\bibitem [{pyq()}]{pyqmc}%
  \BibitemOpen
  \href@noop {} {\enquote {\bibinfo {title} {{PyQMC: Python library for real
  space quantum Monte Carlo}},}\ }\bibinfo {howpublished}
  {\url{https://github.com/WagnerGroup/pyqmc}}\BibitemShut {NoStop}%
\bibitem [{\citenamefont {Sun}\ \emph {et~al.}(2017)\citenamefont {Sun},
  \citenamefont {Berkelbach}, \citenamefont {Blunt}, \citenamefont {Booth},
  \citenamefont {Guo}, \citenamefont {Li}, \citenamefont {Liu}, \citenamefont
  {McClain}, \citenamefont {Sayfutyarova}, \citenamefont {Sharma},
  \citenamefont {Wouters},\ and\ \citenamefont {Chan}}]{pyscf}%
  \BibitemOpen
  \bibfield  {author} {\bibinfo {author} {\bibfnamefont {Q.}~\bibnamefont
  {Sun}}, \bibinfo {author} {\bibfnamefont {T.~C.}\ \bibnamefont {Berkelbach}},
  \bibinfo {author} {\bibfnamefont {N.~S.}\ \bibnamefont {Blunt}}, \bibinfo
  {author} {\bibfnamefont {G.~H.}\ \bibnamefont {Booth}}, \bibinfo {author}
  {\bibfnamefont {S.}~\bibnamefont {Guo}}, \bibinfo {author} {\bibfnamefont
  {Z.}~\bibnamefont {Li}}, \bibinfo {author} {\bibfnamefont {J.}~\bibnamefont
  {Liu}}, \bibinfo {author} {\bibfnamefont {J.~D.}\ \bibnamefont {McClain}},
  \bibinfo {author} {\bibfnamefont {E.~R.}\ \bibnamefont {Sayfutyarova}},
  \bibinfo {author} {\bibfnamefont {S.}~\bibnamefont {Sharma}}, \bibinfo
  {author} {\bibfnamefont {S.}~\bibnamefont {Wouters}}, \ and\ \bibinfo
  {author} {\bibfnamefont {G.~K.~L.}\ \bibnamefont {Chan}},\ }\href {\doibase
  10.1002/wcms.1340} {\bibfield  {journal} {\bibinfo  {journal} {WIREs
  Computational Molecular Science}\ }\textbf {\bibinfo {volume} {8}},\ \bibinfo
  {pages} {e1340} (\bibinfo {year} {2017})}\BibitemShut {NoStop}%
\bibitem [{Wor()}]{Workflow}%
  \BibitemOpen
  \href@noop {} {\enquote {\bibinfo {title} {{S}nakemake workflow and data},}\
  }\bibinfo {howpublished}
  {\url{https://github.com/WagnerGroup/Energy-Entropy/tree/clean_for_paper}}\BibitemShut
  {NoStop}%
\bibitem [{\citenamefont {Ginibre}(1965)}]{Circular_Law}%
  \BibitemOpen
  \bibfield  {author} {\bibinfo {author} {\bibfnamefont {J.}~\bibnamefont
  {Ginibre}},\ }\href {\doibase 10.1063/1.1704292} {\bibfield  {journal}
  {\bibinfo  {journal} {Journal of Mathematical Physics}\ }\textbf {\bibinfo
  {volume} {6}},\ \bibinfo {pages} {440} (\bibinfo {year} {1965})}\BibitemShut
  {NoStop}%
\bibitem [{\citenamefont
  {Mejuto~Zaera}(2021)}]{MejutoZaeraStrongCorrelationThroughSCI2021}%
  \BibitemOpen
  \bibfield  {author} {\bibinfo {author} {\bibfnamefont {C.}~\bibnamefont
  {Mejuto~Zaera}},\ }\emph {\bibinfo {title} {Strong Correlation Through
  Selected Configuration Interaction: From Molecules to Extended Systems}},\
  \href@noop {} {\bibinfo {type} {{PhD} dissertation}},\ \bibinfo  {school}
  {University of California, Berkeley} (\bibinfo {year} {2021})\BibitemShut
  {NoStop}%
\bibitem [{\citenamefont {Dash}\ \emph {et~al.}(2018)\citenamefont {Dash},
  \citenamefont {Moroni}, \citenamefont {Scemama},\ and\ \citenamefont
  {Filippi}}]{dash_perturbatively_2018}%
  \BibitemOpen
  \bibfield  {author} {\bibinfo {author} {\bibfnamefont {M.}~\bibnamefont
  {Dash}}, \bibinfo {author} {\bibfnamefont {S.}~\bibnamefont {Moroni}},
  \bibinfo {author} {\bibfnamefont {A.}~\bibnamefont {Scemama}}, \ and\
  \bibinfo {author} {\bibfnamefont {C.}~\bibnamefont {Filippi}},\ }\href
  {\doibase 10.1021/acs.jctc.8b00393} {\bibfield  {journal} {\bibinfo
  {journal} {Journal of Chemical Theory and Computation}\ }\textbf {\bibinfo
  {volume} {14}},\ \bibinfo {pages} {4176} (\bibinfo {year}
  {2018})}\BibitemShut {NoStop}%
\bibitem [{\citenamefont {Dash}\ \emph {et~al.}(2019)\citenamefont {Dash},
  \citenamefont {Feldt}, \citenamefont {Moroni}, \citenamefont {Scemama},\ and\
  \citenamefont {Filippi}}]{dash_excited_2019}%
  \BibitemOpen
  \bibfield  {author} {\bibinfo {author} {\bibfnamefont {M.}~\bibnamefont
  {Dash}}, \bibinfo {author} {\bibfnamefont {J.}~\bibnamefont {Feldt}},
  \bibinfo {author} {\bibfnamefont {S.}~\bibnamefont {Moroni}}, \bibinfo
  {author} {\bibfnamefont {A.}~\bibnamefont {Scemama}}, \ and\ \bibinfo
  {author} {\bibfnamefont {C.}~\bibnamefont {Filippi}},\ }\href {\doibase
  10.1021/acs.jctc.9b00476} {\bibfield  {journal} {\bibinfo  {journal} {Journal
  of Chemical Theory and Computation}\ }\textbf {\bibinfo {volume} {15}},\
  \bibinfo {pages} {4896} (\bibinfo {year} {2019})}\BibitemShut {NoStop}%
\bibitem [{\citenamefont {Dash}\ \emph {et~al.}(2021)\citenamefont {Dash},
  \citenamefont {Moroni}, \citenamefont {Filippi},\ and\ \citenamefont
  {Scemama}}]{dashTailoringCIPSIExpansions2021}%
  \BibitemOpen
  \bibfield  {author} {\bibinfo {author} {\bibfnamefont {M.}~\bibnamefont
  {Dash}}, \bibinfo {author} {\bibfnamefont {S.}~\bibnamefont {Moroni}},
  \bibinfo {author} {\bibfnamefont {C.}~\bibnamefont {Filippi}}, \ and\
  \bibinfo {author} {\bibfnamefont {A.}~\bibnamefont {Scemama}},\ }\href
  {\doibase 10.1021/acs.jctc.1c00212} {\bibfield  {journal} {\bibinfo
  {journal} {Journal of Chemical Theory and Computation}\ }\textbf {\bibinfo
  {volume} {17}},\ \bibinfo {pages} {3426} (\bibinfo {year}
  {2021})}\BibitemShut {NoStop}%
\bibitem [{\citenamefont {{Morales-Silva}}\ \emph {et~al.}(2021)\citenamefont
  {{Morales-Silva}}, \citenamefont {Jordan}, \citenamefont {Shulenburger},\
  and\ \citenamefont
  {Wagner}}]{morales-silvaFrontiersStochasticElectronic2021}%
  \BibitemOpen
  \bibfield  {author} {\bibinfo {author} {\bibfnamefont {M.~A.}\ \bibnamefont
  {{Morales-Silva}}}, \bibinfo {author} {\bibfnamefont {K.~D.}\ \bibnamefont
  {Jordan}}, \bibinfo {author} {\bibfnamefont {L.}~\bibnamefont
  {Shulenburger}}, \ and\ \bibinfo {author} {\bibfnamefont {L.~K.}\
  \bibnamefont {Wagner}},\ }\href {\doibase 10.1063/5.0053674} {\bibfield
  {journal} {\bibinfo  {journal} {The Journal of Chemical Physics}\ }\textbf
  {\bibinfo {volume} {154}},\ \bibinfo {pages} {170401} (\bibinfo {year}
  {2021})}\BibitemShut {NoStop}%
\bibitem [{\citenamefont {Muechler}\ \emph {et~al.}(2022)\citenamefont
  {Muechler}, \citenamefont {Badrtdinov}, \citenamefont {Hampel}, \citenamefont
  {Cano}, \citenamefont {R\"osner},\ and\ \citenamefont
  {Dreyer}}]{CyrusQuantumEmbeddingMethods2022}%
  \BibitemOpen
  \bibfield  {author} {\bibinfo {author} {\bibfnamefont {L.}~\bibnamefont
  {Muechler}}, \bibinfo {author} {\bibfnamefont {D.~I.}\ \bibnamefont
  {Badrtdinov}}, \bibinfo {author} {\bibfnamefont {A.}~\bibnamefont {Hampel}},
  \bibinfo {author} {\bibfnamefont {J.}~\bibnamefont {Cano}}, \bibinfo {author}
  {\bibfnamefont {M.}~\bibnamefont {R\"osner}}, \ and\ \bibinfo {author}
  {\bibfnamefont {C.~E.}\ \bibnamefont {Dreyer}},\ }\href {\doibase
  10.1103/PhysRevB.105.235104} {\bibfield  {journal} {\bibinfo  {journal}
  {Physical Review B}\ }\textbf {\bibinfo {volume} {105}},\ \bibinfo {pages}
  {235104} (\bibinfo {year} {2022})}\BibitemShut {NoStop}%
\end{thebibliography}%

\appendix 
\setcounter{secnumdepth}{0} 
\section{Appendix: Another quantity to measure correlations: the deviation from idempotence}
\label{appendix:trace object}

We also computed another quantity to measure the correlations of wave functions, as proposed by reference \cite{CyrusQuantumEmbeddingMethods2022}, 
\begin{equation}
    \Lambda = \mathrm{Tr}(\rho-\rho^2).
    \label{eqn:trace_object}
\end{equation}
For an uncorrelated wave function, the 1-RDM is idempotent, i.e. $\rho=\rho^2$, $\Lambda=0$. 
Thus, this quantity can be viewed as a deviation from idempotence.

\begin{figure}
    \centering
    \includegraphics{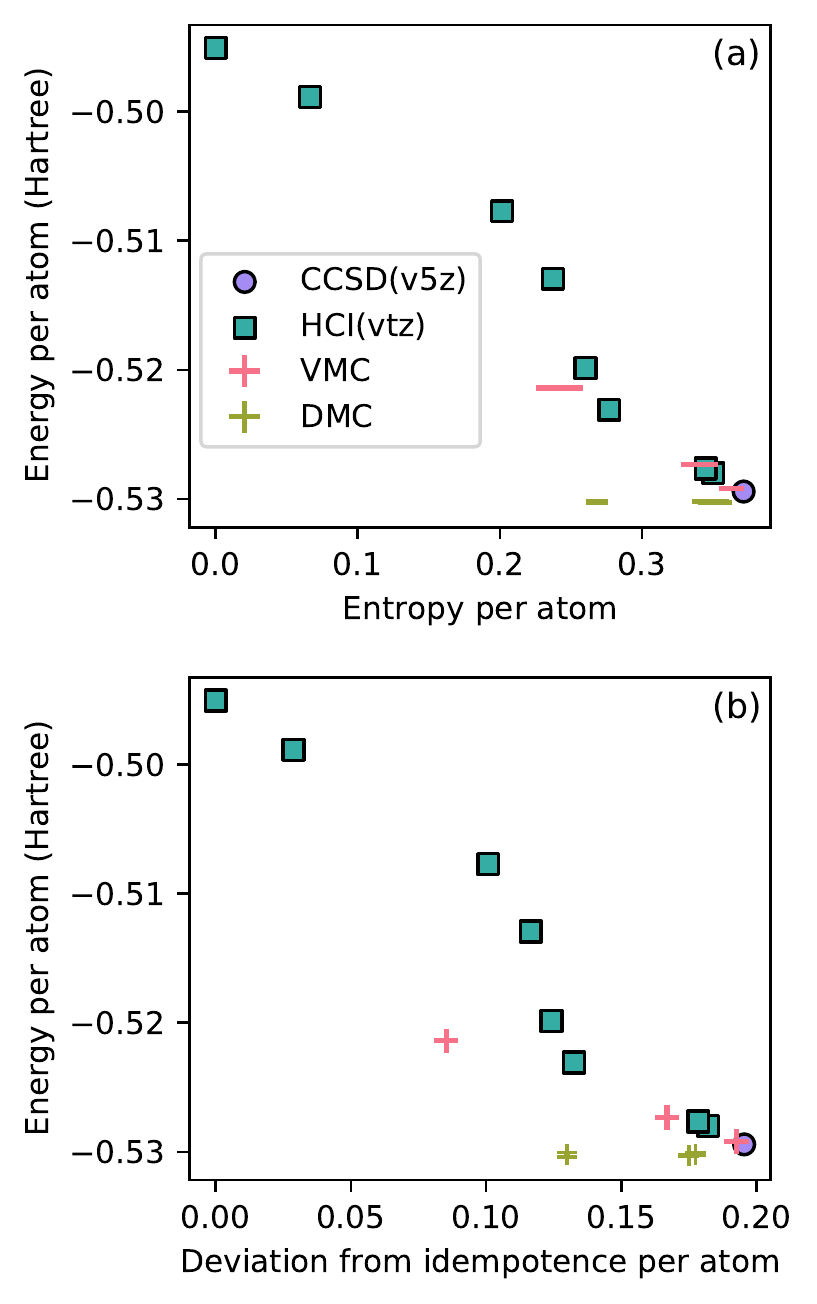}
    \caption{Entropy and the deviation from idempotence show similar dependence on the energy. 
    The calculations were performed in a strongly-correlated system (H6, $r=3.0$ $a_{\text{B}}$), using CCSD (with cc-pV5Z basis), HCI wave functions (with cc-pVTZ basis) using different number of determinants, VMC and FN-DMC using SJ and MSJ trial wave functions. 
    (a) Energy versus entropy per atom. The edges of the bars on the VMC or FN-DMC points represent the lower and upper bounds computed following the method described in section~\ref{sec:circle_reject}. 
    (b) Energy versus the deviation from idempotence per atom. 
    The VMC or FN-DMC points are represented using plus markers. 
    The upper and lower bounds are not evaluated.
    }
    \label{fig: similar dependence of trace object and entropy}
\end{figure}

Fig.~\ref{fig: similar dependence of trace object and entropy} shows that the entropy and the deviation from idempotence $\Lambda$ show similar dependence on the energy.
The shape of the data distribution for the deviation from idempotence is slightly more spread out than that of the entropy, as they have different metrics and units. 
While the deviation from idempotence is easier to evaluate and less susceptible to the stochastic errors in QMC methods, the von Neumann entropy gives more information about correlation through entanglement spectrum, as shown by Fig.~\ref{fig: entanglement spectrum for a strongly correlated system}.

\end{document}